%% file: main.tex
\documentclass[10pt,conference]{IEEEtran}
\IEEEoverridecommandlockouts

\usepackage{algorithmic}
\usepackage{graphicx}
\usepackage{textcomp}
\usepackage{tcolorbox}
\usepackage{pifont}
\usepackage{xcolor}
\usepackage{soul}
\usepackage{booktabs}
\usepackage{multirow}
\usepackage{float}
\usepackage{url}
\usepackage{graphicx}
\usepackage{microtype}
\usepackage{tcolorbox}
\usepackage{enumitem}
\usepackage{seqsplit}
\usepackage[utf8]{inputenc}
\usepackage{newunicodechar,graphicx}
\usepackage[noadjust]{cite}
\usepackage{newunicodechar,graphicx}
\usepackage[hidelinks,bookmarks=true]{hyperref}
\usepackage[switch]{lineno}
\usepackage{enumitem}
\usepackage{seqsplit}
\usepackage{pifont}

\setlist[itemize,enumerate]{noitemsep, topsep=0pt, leftmargin=1.0em}
\setlist{nolistsep}

\hypersetup{
	pdfborder          = {0 0 0},
	breaklinks         = true,
	bookmarksnumbered  = true,
	pdfkeywords        = {Natural Language Processing, Machine Learning, Code Comments, Random Forest, Binary Classification},
	pdftitle           = {Performance Comparison of Binary Machine Learning Classifiers in Identifying Code Comment Types: An Exploratory Study},
	pdfauthor          = {Amila Indika, Peter Y. Washington, Anthony Peruma}
}
\DeclareRobustCommand{\okina}{%
  \raisebox{\dimexpr\fontcharht\font`A-\height}{%
    \scalebox{0.8}{`}%
  }%
}
\newunicodechar{ʻ}{\okina}

\begin{document}

\title{Performance Comparison of Binary Machine Learning Classifiers in Identifying Code Comment Types: An Exploratory Study}


\author{
\IEEEauthorblockN{Amila Indika}
\IEEEauthorblockA{
\textit{University of Hawaiʻi at Mānoa}\\
Honolulu, Hawaiʻi, USA \\
amilaind@hawaii.edu}
\and
\IEEEauthorblockN{Peter Y. Washington}
\IEEEauthorblockA{
\textit{University of Hawaiʻi at Mānoa}\\
Honolulu, Hawaiʻi, USA \\
pyw@hawii.edu}
\and
\IEEEauthorblockN{Anthony Peruma}
\IEEEauthorblockA{
\textit{University of Hawaiʻi at Mānoa}\\
Honolulu, Hawaiʻi, USA \\
peruma@hawaii.edu}

}




\maketitle

\begin{abstract}
Code comments are vital to source code as they help developers with program comprehension tasks. Written in natural language (usually English), code comments convey a variety of different information, which are grouped into specific categories. In this study, we construct 19 binary machine learning classifiers for code comment categories that belong to three different programming languages. We present a comparison of performance scores for different types of machine learning classifiers and show that the Linear SVC classifier has the highest average F1 score of 0.5474.
\end{abstract}


\section{Introduction}
Program comprehension is an essential activity in any software maintenance task \cite{rajlich2002role}, which involves developers reading the code to understand its purpose, behavior, and shortcomings (if any) \cite{maalej2014comprehension}. However, while source code is helpful, it is constrained by the semantics and syntax of programming language in the amount of information it can provide. In contrast, code comments are written in natural language (such as English) and are not restricted to any specific syntax or grammar; they can provide information that might not be immediately obvious from the source code alone. Developers use code comments to document various aspects of their code, from explaining behavior to documenting workarounds to including metadata (such as a license), all expressed in natural language \cite{Pascarella2017MSR, rani2021}. While such information in comments is useful, it would be more productive for developers to group such comments into related categories automatically. This would help developers locate pertinent information faster, such as improving code quality by removing/replacing deprecated code.

The goal of this study is to understand the \textit{extent to which different types of binary machine learning classifiers can identify code comment types}. Our findings show that while no single classifier single-handedly achieves the highest F1 score for each individual comment type, the \textbf{Linear SVC has the highest average F1 score of 0.5474}. The outcome of this study will help the research community understand the strengths and shortcomings of using such models for this particular task and discover avenues for future research in this area.

\begin{figure*}[t]
 	\centering
 	\includegraphics[trim=0cm 0cm 0cm 0cm,scale=0.75]{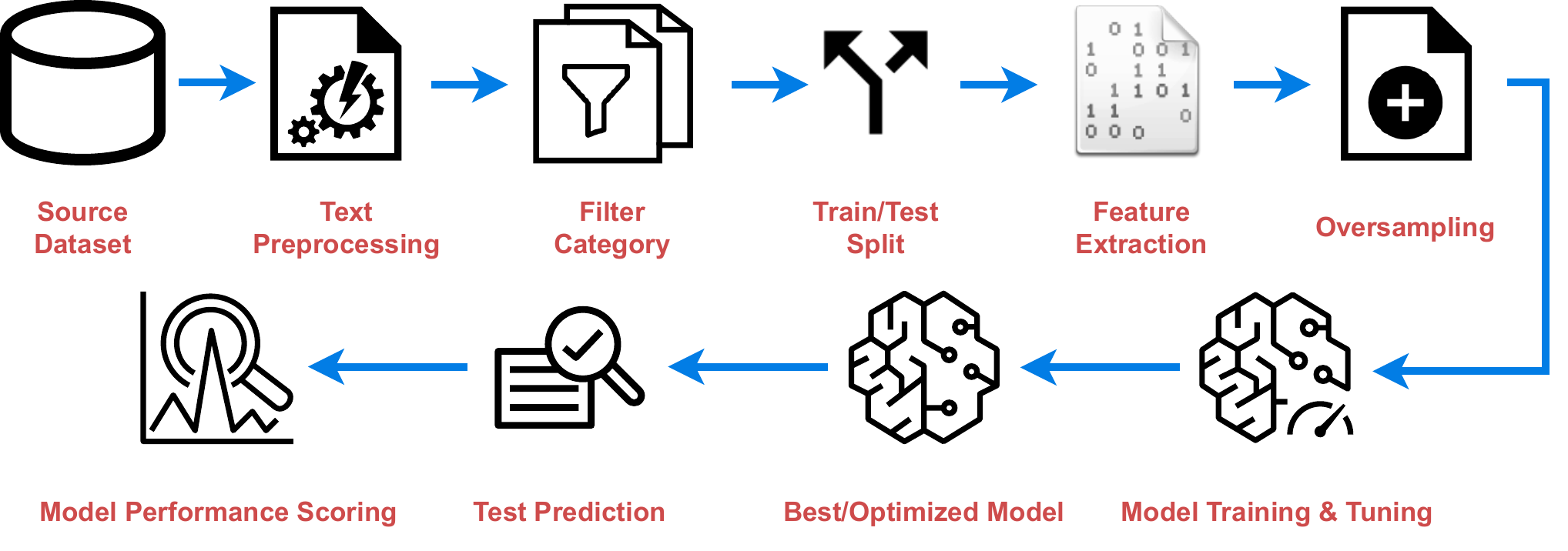}
 	\caption{Overview of our methodology for constructing the comment classification models.}\vspace{-2mm}
 	\label{Figure:diagram_experiment}
\end{figure*}

\input{Table_Comparison.tex}

\section{Experiment Design}
\label{Section:experiment_design}
In this section, we discuss the methodology of our study. Figure \ref{Figure:diagram_experiment} depicts an outline of the activities in our experiment methodology, which we describe below. Furthermore, our dataset and code are available for replication/extension at \cite{ProjectWebsite}.

\subsection{Source Dataset}
We utilize an existing dataset of 6,738 class comment sentences produced by Rani et al. \cite{rani2021} and made available as part of the NLBSE 2023 tool competition \cite{nlbse2023}. These comment sentences are obtained from 20 open-source projects implemented using Java, Python, and Pharo. The authors of the dataset also perform text preprocessing and splitting of the comments. Furthermore, the dataset also indicates the category (i.e., class) associated with the comment sentence. Below is a breakdown of comments categories: 
\begin{itemize}
    \item Java - 2418 comment sentences in 7 categories: \textit{summary}, \textit{pointer}, \textit{deprecation}, \textit{rational}, \textit{ownership}, \textit{usage}, and \textit{expand}
    \item Pharo - 1765 comment sentences in 7 categories: \textit{key messages}, \textit{intent}, \textit{class references}, \textit{example}, \textit{key implementation points}, \textit{responsibilities}, and \textit{collaborators}
    \item Python - 2555 comment sentences in 5 categories: \textit{summary}, \textit{parameters}, \textit{usage}, \textit{development notes}, and \textit{expand}
\end{itemize}

\noindent For each programming language in the dataset, we execute the activities in the subsequent subsections.

\subsection{Text Preprocessing}
Since code comments are crafted using natural language, there exists the possibility of developers composing these comments using a variety of tokens that include numeric digits and non-alphanumeric characters (such as brackets, parentheses, periods, etc.). Therefore, to improve model classification performance, it is essential to transform the comment sentences to a standard and convenient format (i.e., text normalization) \cite{jurafsky2019speech}. We achieve text normalization through a series of processing activities: \\

\begin{itemize}
\item Removal of whitespaces from the start and end of the text
\item Expansion of contractions (e.g., `I'm' $\rightarrow$ `I am')
\item Removal of non-alphanumeric characters
\item Removal of single-character words
\item Convert text to lowercase
\item Removal of stopwords (using the NLTK stopword list \cite{bird2004nltk})
\item Reduce words to their base form or root using stemming
\item Replace numeric digits with a NUM token
\item Replace empty string with an EMT token
\end{itemize}

\subsection{Filter Category}
The purpose of this study is the construction of binary classifiers for each of the 19 categories mentioned above. Therefore, for each programming language dataset, we filter the comments for each category. For example, when processing the Java comments, we filter on each of the 7 categories. For the activities in the subsequent subsections, we utilize this subset of comments (i.e., each programming language category) to build and evaluate the classification model.

\subsection{Train/Test Split}
As part of measuring the model's effectiveness in correctly classifying comment sentences, it must be tested against a held-out test set containing comment sentences with known classes. We split the filtered dataset into a training and test dataset. Contained within the source dataset is a column that indicates if a specific comment sentence belongs to the test or training set; we use this column to create the train/test datasets.

\subsection{Feature Representation}
The raw comment text must be converted into a set of numerical features that can be used as input to the model. Most machine learning algorithms can only process numerical data instead of raw text. Furthermore, converting text to numerical data reduces the complexity of the data. To this extent, we utilize the Term Frequency-Inverse Document Frequency (TF-IDF) approach of representing the raw comment sentence text as numerical values. TF-IDF calculates the importance of each word in a document based on its frequency and rarity, which we then use as features for the machine learning model \cite{manning2008introduction}. In this study, we only utilize the comment sentence text as input to the classification model. In our process, we first fit the vectorizer on the training data and transform the training data into a matrix of TF-IDF features. Finally, we utilize this generated vocabulary and TF-IDF representation scheme to transform the test data into a matrix of TF-IDF features. This approach ensures that the resulting test data TF-IDF matrix has the same number of columns (i.e., the same vocabulary) as the training data and avoids model performance issues.

\subsection{Oversampling}
We applied an oversampling technique to the training datasets to mitigate the issue of having imbalanced classes. Having an imbalanced dataset will lead to a bias in the model towards the majority class due to the disproportionate amount of data from that class. For this study, we utilized random oversampling to generate new samples for the under-represented classes in the training dataset \cite{amr2020hands}. We do not oversample the test dataset, as we evaluate the model using metrics that inherently account for lack of balance (precision, recall, and F1 score).

\subsection{Model Training \& Tuning}
Given the complexity of various machine learning classification algorithms, selecting the most optimal algorithm for a given problem using only theoretical arguments can be challenging. Hence, in this study, we evaluate 8 common machine learning classification algorithms frequently utilized in software engineering research \cite{Azeem2019ML,sharma2021survey, AlOmar2021ML}. These classification algorithms fall into different categories: \\

\begin{itemize}
    \item \textbf{Naive Bayes (NB)}: \textit{Multinomial NB} and \textit{Bernoulli NB}
    \item \textbf{Support Vector Machines}: \textit{Linear Support Vector Classifier}
    \item \textbf{Trees}: \textit{Decision Tree} and \textit{Random Forest}
    \item \textbf{Nearest Neighbors}: \textit{K-Nearest Neighbors}
    \item \textbf{Linear Model}: \textit{Logistic Regression}
    \item \textbf{Neural Network}: \textit{Multi-Layer Perceptron} \\
\end{itemize} 

For each classification algorithm listed above, we use grid search in conjunction with a 10-fold cross-validation to perform a search over specified hyperparameter values to evaluate the model's performance. Grid search utilizes a brute force technique to evaluate all combinations of hyperparameters to obtain the best performance \cite{dangeti2017statistics}. We tuned at least one hyperparameter associated with each classifier in our list. The hyperparameter values were either a range of values or a predefined set of values. For example, we evaluated the performance of the Decision Tree classifier by changing the \textit{maximum depth} hyperparameter from 5 to 100 in increments of 5. The complete set of evaluated hyperparameters and their range of possible values are available in the source code contained within the replication package (available at \cite{ProjectWebsite}).

\subsection{Best/Optimized Model}
After hyperparameter tuning, we obtain a model for each classification algorithm that performs the best when built on the training data. For each of these optimum models, we capture the values of the hyperparameters and save the model. 

\subsection{Test Predictions}
We utilize the optimum model for each classification algorithm to predict the test data values (i.e., unseen data). We save the predictions as a CSV file.

\subsection{Model Performance Scoring}
Next, we utilize standard balanced binary classification metrics (precision, recall, and F1 score) to evaluate the optimum model's performance in predicting the test data values. To calculate these scores, we utilize the count of true positive (TP), false positive (FP), and false negative (FN) instances in the predicted test data values. Below we list the formulas for the three performance measures:

\[
    Precision = \frac{TP}{TP+FP}
\]

\[
    Recall = \frac{TP}{TP+FN}
\]    

\[    
    F1 = 2*\frac{Precision*Recall}{Precision+Recall}
\]

We calculate the precision, recall, and F1 scores for each classification model for each programming language category. Since we are evaluating eight classifiers across 19 categories, we calculate 190 instances of precision, recall, and F1 scores. While there are different variants of the F1-score, such as `micro', `macro', etc., we utilize the `binary' option since our problem is a binary classification task \cite{ciaburro2019python,sklearn:f1}. Using another variant will give different (and possibly higher) scores.

\section{Experiment Results}
\label{Section:experiment_results}

\input{Table_LinearSVC_score.tex}

In this section, we report on the results of our experiment. Table \ref{Table:comparison_score} shows the F1 scores we obtain for each classification model for each programming language category. From this table, we observe that no single classification model achieves the highest F1 score for each programming language category. Looking at the average F1 scores, we observe that \textbf{Linear SVC has the highest average F1 score of 0.5474}. It is interesting to note that all models were able to successfully classify all test data instances of the Java Ownership category, 489 instances. We also observe that the models seem to fare well with Java comments than with Pharo and Python comments. In Table \ref{Table:scores}, we provide the precision and recall score of the optimal Linear Support Vector Classifier for each programming language category.


It is interesting to note that while several machine learning methods with varying complexities were explored, the two best-performing models were constrained to learning linearly separating boundaries: logistic regression and a support vector classifier with a linear kernel. These methods are less prone to overfitting due to the simplicity of the decision boundary.

\section{Threats To Validity}
\label{Section:threats}
Even though our study includes a variety of classification algorithms, other types of classifiers can be utilized on the same dataset. Likewise, there is a possibility that alternative preprocessing and oversampling techniques, features (for instance, using the NEON tool \cite{DiSorboVPCP21}), and hyperparameters would yield improved performance scores. 

\section{Conclusion \& Future Work}
\label{Section:conclusion}
In this study, we explored the performance of different types of machine learning classifiers in performing binary classification of code comment sentences. Our results show that even though no single classifier can achieve high F1 scores for all code comment categories, the simplest model we developed (i.e., a Linear SVC), which is constrained to linear decision boundaries, yielded the highest average F1 score. 

We did not observe a noticeable consistent pattern which might provide potential explanations as to why the best classifier performs better for some categories than for others. Understanding these discrepancies is a salient opportunity for further evaluation.

Future work includes evaluating attention-based deep learning models, such as BERT \cite{devlin-etal-2019-bert} and GPT-3 \cite{brown2020language}. We hypothesize that such large pre-trained language models can outperform the models presented here, although the extent to which is to be determined due to the heterogeneous nature of the prediction task. It would also be interesting to explore the feature vector representation space beyond TF-IDF, such as automatically learned representations (notably, Word2Vec \cite{mikolov2013efficient, mikolov2013distributed}).

Our source code (along with instructions) and generated optimal models are available for download at \cite{ProjectWebsite}.

\section{Acknowledgements}
\label{Section:acknowledgements}
The technical support and advanced computing resources from University of  Hawaiʻi Information Technology Services – Cyberinfrastructure, funded in part by the National Science Foundation CC* awards \#2201428 and \#2232862 are gratefully acknowledged.

\bibliographystyle{ieeetr}
\bibliography{main}

\end{document}

%% file: Table_Comparison.tex
\begin{table*}
\centering
\caption{Comparison of F1 scores associated with each machine learning classifier for each category. The score in green represents the highest F1 score for the specific category. The score in blue is the classifier that has the highest average F1 score. }\vspace{-2mm}
\label{Table:comparison_score}
\begin{tabular}{@{}lrrrrrrrr@{}}
\toprule
\multicolumn{1}{c}{\textbf{Category}} & \multicolumn{1}{c}{\textbf{\begin{tabular}[c]{@{}c@{}}Decision\\ Tree\end{tabular}}} & \multicolumn{1}{c}{\textbf{\begin{tabular}[c]{@{}c@{}}K-Nearest\\ Neighbors\end{tabular}}} & \multicolumn{1}{c}{\textbf{\begin{tabular}[c]{@{}c@{}}Bernoulli\\ Naive Bayes\end{tabular}}} & \multicolumn{1}{l}{\textbf{\begin{tabular}[c]{@{}l@{}}Multi-Layer\\ Perceptron\end{tabular}}} & \multicolumn{1}{c}{\textbf{\begin{tabular}[c]{@{}c@{}}Multinomial\\ Naive Bayes\end{tabular}}} & \multicolumn{1}{c}{\textbf{\begin{tabular}[c]{@{}c@{}}Random\\ Forest\end{tabular}}} & \multicolumn{1}{c}{\textbf{\begin{tabular}[c]{@{}c@{}}Logistic\\ Regression\end{tabular}}} & \multicolumn{1}{c}{\textbf{\begin{tabular}[c]{@{}c@{}}Linear\\ SVC\end{tabular}}} \\ \midrule
\multicolumn{9}{c}{\textit{\textbf{JAVA}}}                                                                                                                                                                                                                                                                                                                                                                                                                                                                                                                                                                                                                                                                                                                                                        \\
Deprecation                           & 0.5797                                                                               & {\color[HTML]{34A853} \textit{\textbf{0.8163}}}                                            & 0.5195                                                                                       & 0.7600                                                                                        & 0.6032                                                                                         & 0.8085                                                                               & 0.7600                                                                                     & 0.7170                                                                            \\
Pointer                               & 0.6069                                                                               & 0.5890                                                                                     & 0.5697                                                                                       & 0.5556                                                                                        & 0.6301                                                                                         & {\color[HTML]{34A853} \textit{\textbf{0.6418}}}                                      & 0.6265                                                                                     & \textit{0.5850}                                                                   \\
Summary                               & 0.5787                                                                               & 0.5030                                                                                     & {\color[HTML]{34A853} \textit{\textbf{0.6327}}}                                              & 0.5532                                                                                        & 0.5905                                                                                         & 0.5455                                                                               & 0.5842                                                                                     & 0.5684                                                                            \\
Expand                                & 0.4857                                                                               & 0.5833                                                                                     & 0.5704                                                                                       & 0.5970                                                                                        & 0.6097                                                                                         & 0.6160                                                                               & {\color[HTML]{34A853} \textit{\textbf{0.6264}}}                                            & 0.6100                                                                            \\
Ownership                             & {\color[HTML]{34A853} \textit{\textbf{1.0000}}}                                      & {\color[HTML]{34A853} \textit{\textbf{1.0000}}}                                            & {\color[HTML]{34A853} \textit{\textbf{1.0000}}}                                              & {\color[HTML]{34A853} \textit{\textbf{1.0000}}}                                               & {\color[HTML]{34A853} \textit{\textbf{1.0000}}}                                                & {\color[HTML]{34A853} \textit{\textbf{1.0000}}}                                      & {\color[HTML]{34A853} \textit{\textbf{1.0000}}}                                            & {\color[HTML]{34A853} \textit{\textbf{1.0000}}}                                   \\
Rational                              & 0.5455                                                                               & 0.5517                                                                                     & 0.5854                                                                                       & 0.6122                                                                                        & 0.5669                                                                                         & {\color[HTML]{34A853} \textit{\textbf{0.6667}}}                                      & 0.5833                                                                                     & \textit{0.5920}                                                                   \\
Usage                                 & 0.5931                                                                               & 0.7255                                                                                     & 0.7612                                                                                       & 0.7493                                                                                        & {\color[HTML]{34A853} \textit{\textbf{0.7740}}}                                                & 0.7435                                                                               & 0.7507                                                                                     & 0.7413                                                                            \\ \midrule
\multicolumn{9}{c}{\textit{\textbf{PHARO}}}                                                                                                                                                                                                                                                                                                                                                                                                                                                                                                                                                                                                                                                                                                                                                       \\
Class References                      & 0.1778                                                                               & 0.2963                                                                                     & 0.2353                                                                                       & 0.1600                                                                                        & 0.2105                                                                                         & {\color[HTML]{34A853} \textit{\textbf{0.3000}}}                                      & 0.2069                                                                                     & 0.2857                                                                            \\
Example                               & 0.7132                                                                               & 0.7320                                                                                     & {\color[HTML]{34A853} \textit{\textbf{0.8063}}}                                              & 0.7525                                                                                        & 0.7778                                                                                         & 0.7438                                                                               & 0.7509                                                                                     & 0.7603                                                                            \\
Key Implementation Points             & 0.2642                                                                               & 0.3125                                                                                     & 0.3529                                                                                       & 0.3571                                                                                        & 0.3902                                                                                         & 0.2424                                                                               & 0.3191                                                                                     & {\color[HTML]{34A853} \textit{\textbf{0.4124}}}                                   \\
Collaborators                         & 0.3810                                                                               & 0.2979                                                                                     & 0.2712                                                                                       & 0.2609                                                                                        & 0.2308                                                                                         & 0.1714                                                                               & 0.4138                                                                                     & {\color[HTML]{34A853} \textit{\textbf{0.4231}}}                                   \\
Intent                                & 0.4854                                                                               & 0.3488                                                                                     & 0.3208                                                                                       & 0.2759                                                                                        & 0.3390                                                                                         & {\color[HTML]{34A853} \textit{\textbf{0.5479}}}                                      & 0.4554                                                                                     & 0.4301                                                                            \\
Key Messages                          & 0.4628                                                                               & 0.4407                                                                                     & 0.5733                                                                                       & {\color[HTML]{34A853} \textit{\textbf{0.6055}}}                                               & {\color[HTML]{34A853} \textit{\textbf{0.5797}}}                                                & 0.5149                                                                               & 0.5                                                                                        & 0.5254                                                                            \\
Responsibilities                      & 0.3407                                                                               & 0.3193                                                                                     & 0.3867                                                                                       & 0.4192                                                                                        & 0.3617                                                                                         & 0.3585                                                                               & {\color[HTML]{34A853} \textit{\textbf{0.4485}}}                                            & 0.4218                                                                            \\ \midrule
\multicolumn{9}{c}{\textit{\textbf{PYTHON}}}                                                                                                                                                                                                                                                                                                                                                                                                                                                                                                                                                                                                                                                                                                                                                      \\
Development Notes                     & {\color[HTML]{34A853} \textit{\textbf{0.3472}}}                                      & 0.2314                                                                                     & 0.2995                                                                                       & 0.3066                                                                                        & 0.2769                                                                                         & 0.3333                                                                               & 0.2838                                                                                     & 0.3145                                                                            \\
Parameters                            & 0.6523                                                                               & 0.5831                                                                                     & 0.6359                                                                                       & 0.6203                                                                                        & 0.6387                                                                                         & 0.6605                                                                               & 0.6565                                                                                     & 0.6337                                                                            \\
Summary                               & 0.3169                                                                               & 0.3444                                                                                     & 0.4017                                                                                       & 0.3579                                                                                        & 0.4050                                                                                         & 0.3432                                                                               & {\color[HTML]{34A853} \textit{\textbf{0.4138}}}                                            & 0.3774                                                                            \\
Expand                                & 0.3514                                                                               & 0.4599                                                                                     & 0.4571                                                                                       & {\color[HTML]{34A853} \textit{\textbf{0.4643}}}                                               & 0.4534                                                                                         & 0.4293                                                                               & 0.4519                                                                                     & {\color[HTML]{34A853} \textit{\textbf{0.4636}}}                                   \\
Usage                                 & 0.4859                                                                               & 0.4266                                                                                     & 0.5481                                                                                       & 0.5233                                                                                        & 0.5342                                                                                         & 0.5275                                                                               & 0.5519                                                                                     & 0.5387                                                                            \\ \midrule
\textit{\textbf{AVERGE F1 SCORE}}     & 0.4931                                                                               & 0.5033                                                                                     & 0.5225                                                                                       & 0.5227                                                                                        & 0.5249                                                                                         & {\color[HTML]{333333} 0.5366}                                                        & {\color[HTML]{333333} 0.5465}                                                              & {\color[HTML]{0000FF} \textit{\textbf{0.5474}}}                                   \\ \bottomrule
\end{tabular}\vspace{-3mm}
\end{table*}

%% file: Table_LinearSVC_score.tex
\begin{table}
\centering
\caption{Breakdown of the performance scores of the \textbf{{linear svc}} classifiers for each programming language category.}\vspace{-2mm}
\label{Table:scores}
\begin{tabular}{@{}lrrr@{}}
\toprule
\multicolumn{1}{c}{\textbf{Category}}               & \multicolumn{1}{c}{\textbf{Precision}} & \multicolumn{1}{c}{\textbf{Recall}} & \multicolumn{1}{c}{\textbf{F1}} \\ \midrule
\multicolumn{4}{c}{\textit{\textbf{JAVA}}}                                                                                                                           \\
Deprecation                                         & 0.7308                                 & 0.7037                              & 0.7170                          \\
Pointer                                             & 0.5972                                 & 0.5733                              & 0.5850                          \\
Summary                                             & 0.5243                                 & 0.6207                              & 0.5684                          \\
Expand                                              & 0.5985                                 & 0.6220                              & 0.6100                          \\
Ownership                                           & 1.0000                                 & 1.0000                              & 1.0000                          \\
Rational                                            & 0.5441                                 & 0.6491                              & 0.5920                          \\
Usage                                               & 0.7277                                 & 0.7554                              & 0.7413                          \\ \midrule
\multicolumn{4}{c}{\textit{\textbf{PHARO}}}                                                                                                                          \\
Class References                                    & 0.3636                                 & 0.2353                              & 0.2857                          \\
Example                                             & 0.7929                                 & 0.7303                              & 0.7603                          \\
Key Implementation Points                           & 0.4082                                 & 0.4167                              & 0.4124                          \\
Collaborators                                       & 0.4583                                 & 0.3929                              & 0.4231                          \\
Intent                                              & 0.4167                                 & 0.4444                              & 0.4301                          \\
Key Messages                                        & 0.5636                                 & 0.4921                              & 0.5254                          \\
Responsibilities                                    & 0.3974                                 & 0.4493                              & 0.4218                          \\ \midrule
\multicolumn{4}{c}{\textit{\textbf{PYTHON}}}                                                                                                                         \\
Development Notes                                   & 0.2660                                 & 0.3846                              & 0.3145                          \\
Parameters                                          & 0.5956                                 & 0.6770                              & 0.6337                          \\
Summary                                             & 0.3361                                 & 0.4301                              & 0.3774                          \\
Expand                                              & 0.4322                                 & 0.5000                              & 0.4636                          \\
Usage                                               & 0.5438                                 & 0.5337                              & 0.5387                          \\ \midrule
\multicolumn{1}{r}{\textit{\textbf{AVERAGE SCORE}}} & \textbf{0.5419}                        & \textbf{0.5585}                     & \textbf{0.5474}                 \\ \bottomrule
\end{tabular}\vspace{-3mm}
\end{table}